# Facial and meridional isomers of tris(bidentate) Ir(III) complexes: unravelling their different excited state reactivity

Sylvio Arroliga-Rocha,[a] Daniel Escudero[a,]*

[a] Chimie Et Interdisciplinarité, Synthèse, Analyse, Modélisation (CEISAM), UMR CNRS no. 6320, BP 92208, Université de Nantes, 2, Rue de la Houssinière, 44322 Nantes, Cedex 3, France.

Supporting Information Placeholder

**ABSTRACT:** The use of tris(bidentate) Ir(III) complexes as light active components in phosphorescent organic light-emitting diodes (PhOLEDs) is currently the state-of-the-art technology to attain long-lasting and highly-performant devices. Still, further improvements of their operational lifetimes are required for their practical use in lighting and displays. Facial/meridional stereoisomerism of the tris(bidentate) Ir(III) architectures strongly influences their emissive properties and thereto their PhOLEDs performances and operational device stabilities. This work underpins at the first principles level the different excited state reactivity of facial and meridional stereoisomers of a series of tris(bidentate) Ir(III) complexes, which is found to be originated on the presence of distinct $^3$MC deactivation pathways. These deactivation pathways are herein presented for the first time for the meridional isomers. Finally, we propose some phosphor design strategies.

## Introduction

Phosphorescent-based OLEDs (PhOLEDs) are the most widespread electroluminescent devices due to their record external quantum efficiencies (EQE) and low power consumption.[1] In these devices, electroluminescence originates from triplet excitons localized at the dopant materials, i.e., typically organometallic complexes of Ir(III) and/or Pt(II),[2] which promote very efficient intersystem crossing and phosphorescence processes, and thus guaranteeing that internal quantum efficiencies (IQE) of up to 100% can be attained. For practical applications, besides large IQE values, it is indispensable that the devices possess high operational stability.[3] Intrinsic chemical degradation of both host and dopant molecules upon PhOLEDs' continuous operation is responsible of harming the device operational lifetimes.[4] Deterioration of PhOLEDs devices leads to a long-term decrease of the overall luminance,[5] since the chemically degraded products act as nonradiative recombination sites, charge traps and/or luminescence quenchers. In recent years, many experimental[6] and computational[7,8] efforts have been devoted to unveil possible intrinsic degradation mechanisms for both host and dopant molecules. The degradation routes of PhOLEDs materials root on parasitic annihilation processes between excited states (i.e., exciton-polaron and/or exciton-exciton) that lead to the formation of hot excitons (that is, highly energetic excited states of up to ≅6.0 eV).[9] Whilst exciton-polaron annihilation[10] but also radical ion pair annihilation[11] have been highlighted as highly relevant mechanism for the degradation of both host and dopant materials, triplet-triplet annihilation (TTA) is of uttermost relevance for Ir(III) dopants.[12] Some of us recently reported ligand dissociation reactions arising from hot excitons in these complexes.[7] We found that both kinetic and thermodynamic aspects controlling the formation of the triplet metal-centered ($^3$MC) excited states do play an important role in these degradation processes. Furthermore, the thermal population of $^3$MC states is also responsible for the diminished radiative efficiency of many of these complexes at room temperature; especially for blue phosphors.[13,14,15] In a nutshell, one should propose tailored phosphor architectures to avoid the population of $^3$MC states, since they ultimately compromise both PhOLEDs' efficiencies and stabilities.

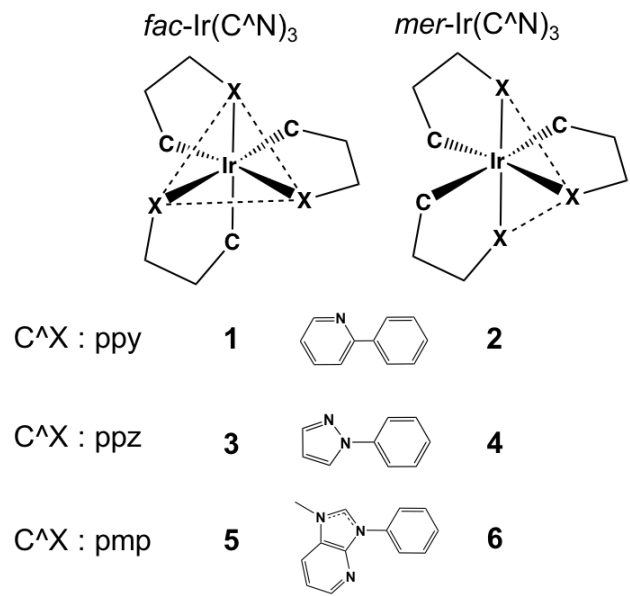

Scheme 1. Chemical structure of complexes 1-6.

To date, little attention has been paid to the excited state properties of the different isomers of tris(bidentate) Ir(III) complexes.[16,17] In Scheme 1 the facial (*fac*-) and meriodinal (*mer*-) stereoisomers are depicted. Research interests have mainly been focused on the *fac*-isomers, since they usually correspond to the most thermodynamically stable products, and thus they are easier to synthesize. Both thermal and photochemical *mer*-to-*fac* isomerization is generally observed for tris(bidentate) Ir(III) complexes.[17] The *fac*-isomers generally possess larger photoluminescence quantum yields than their

*mer*-counterparts.[17] This is so because larger nonradiative decay rates are generally obtained for the latter isomers, whilst the radiative rates remain alike for both isomers. Additionally, the *mer*-isomer is usually more susceptible to degradation than the *fac*-isomer upon PhOLED operation.[18] These features hold generally true for all kind of cyclometalating ligands (e.g., phenylpyridine (ppy) and phenylpyrazole (ppz), see Scheme 1).[19] Although, the presence of different nonradiative channels involving several $^3$MC states was postulated as a likely origin of the different excited state reactivity between the stereoisomers,[19] no compelling proof of it has been provided yet. Conversely, in the case of *N*-heterocyclic carbene (NHC) ligands, such as e.g., the tris-(*N*-phenyl, *N*-methyl-pyridoimidazol-2-yl (pmp) shown in Scheme 1, both *mer*- and *fac*-isomers possess comparable radiative efficiencies and PhOLEDs' operational stabilities.[20] In this contribution we explore the excited state potential energy surfaces (PES) of a series of *fac*- and *mer*-tris(bidentate) Ir(III) complexes, with the final aim of underpinning their distinct photoreactivity that strongly influences their PhOLEDs' operational stability. We finally provide some phosphor design rules.

**Results and Discussion**

Scheme 1 gathers the tris(bidentate) Ir(III) complexes studied here (**1-6**). Experimentally, the photophysical properties of both the *fac*- and mer-isomers of Ir(ppy)$_3$, Ir(ppz)$_3$ and Ir(pmp)$_3$ were studied in detail.[17] The experimental photoluminescence quantum yields of **1-6** are collected in Table S1. In addition, PhOLEDs devices with both stereoisomers of Ir(ppz)$_3$ and Ir(pmp)$_3$ were built up and their performances and degradation pathways were analyzed.[18,20] From a computational viewpoint, most of the available work to date is based on the *fac*-isomers,[15] whilst the *mer*-isomers remain largely unexplored. Photoluminescence in tris(bidentate) Ir(III) complexes arises from a triplet excited state of predominant metal-to-ligand charge transfer ($^3$MLCT) character.[2] The relevant geometrical parameters of the optimal $^3$MLCT geometries of complexes 1-6 are found in Figure S1. The radiative deactivation ($k_r$) from the $^3$MLCT state directly competes with nonradiative decay from from this state ($k_{ISC}$). Additionally, $^3$MC states can be thermally populated from the $^3$MLCT state.[13] $^3$MC states are highly distorted with respect to the $^3$MLCT and ground state ($^1$GS) geometries, and they are known to effectively suppress photoluminescence through minimum energy crossing points (MECP) with the $^1$GS PES.[14] In the common C^N cyclometalated complexes (**1-4**), the pseudo-octahedral disposition at the $^3$MLCT and $^1$GS geometries is distorted to attain a pentacoordinate trigonal bipyramid one at the $^3$MC geometry, displaying one broken Ir-N bond. Note that Ir-N bonds are more prone to dissociation than their Ir-C counterparts, as the latter are generally stronger. As recently shown by one of us, the global photoluminescence decay kinetics cannot be reproduced without tacking into account the $^3$MLCT→$^3$MC→$^1$GS channel in the simulations, and thus further corroborating the important role of the $^3$MC state in the nonradiative decay mechanisms.[21] In the following we deeply explore the triplet excited state PES of **1-6**, with the goal of describing novel $^3$MC deactivation pathways and their role in the excited state reactivity. DFT optimizations of the relevant stationary points along these pathways (ground and excited state minima, transition states (TS) and MECP) were performed (see Computational Details), as this method has been shown to succeed in reaching a continuous adiabatic description of the lowest triplet PES.[14]

***fac*-/*mer*-Ir(ppy)$_3$ (1/2):** The excited state PES of *fac*-Ir(ppy)$_3$ have been largely explored.[14,21] Up to date, only one type of $^3$MC state was found for this complex. The most relevant geometrical features of this $^3$MC state, along with its singly occupied natural transition orbitals (NTO), are shown in Figure 1a-b, respectively. The $^3$MC minimum displays the elongation and fully rupture of one axial Ir-N bond (3.39 Å, see Figure 1a). The NTO analysis indicates that this $^3$MC state involves the $d_{x^2-y^2}$-like $e_g$ orbital (see LUMO in Figure 1b). Herein, we performed further attempts to optimize other $^3$MC states of **1** by manually mapping the topology of its lowest adiabatic triplet PES. These attempts were unsuccessful, and thus, we believe that the $^3$MC deactivation channel shown in Figure 1c is the only one available for *fac*-Ir(ppy)$_3$. Such a channel possesses a limiting barrier of ca. 10 kcal/mol according to our PWPB95 calculations (see Figure 1c and Table 1). The computed profile is very shallow until reaching the MECP. As corroborated through the computed IRC profile, once the MECP is surpassed the system recovers the $^1$GS geometry without chemical transformation. As it was previously found for **1**, this channel is only relevant at temperatures above 298 K.[13,21]

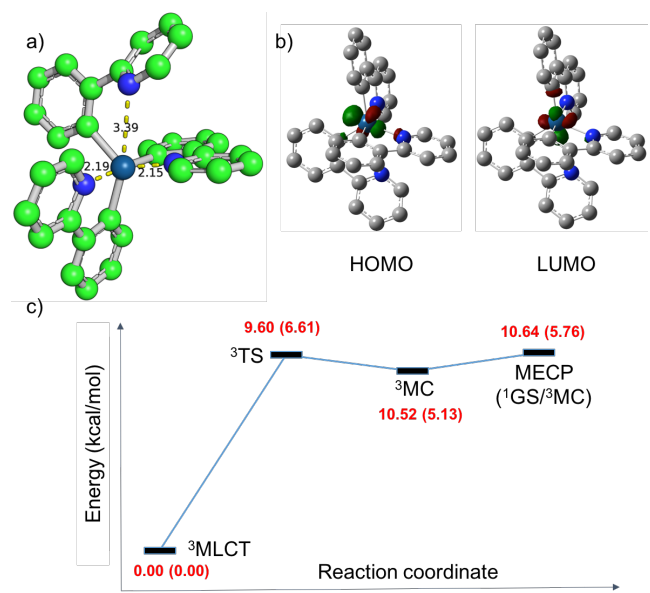

**Figure 1.** a) Relevant geometrical parameters at the $^3$MC optimized geometry of 1. b) Computed HOMO and LUMO NTOs at the $^3$MC optimized geometry. c) PWPB95/6-31G* (B3LYP values between parentheses) energetic profile of the temperature-dependent nonradiative pathway of 1. The reference is the $^3$MLCT emissive state.

Let's now turn the discussion to *mer*-Ir(ppy)$_3$ (**2**). Its $^3$MC deactivation pathways have not been disclosed yet. In contrast to **1**, this complex possesses two Ir-N bonds in *trans* position (see Scheme 1). Such a disposition may open up the appearance of additional $^3$MC states, as e.g., the bonding/antibonding interactions between the $d_{z^2}$-like $e_g$ orbital and the two nitrogen lone pairs can easily be modified by shortening/enlarging the weaker Ir-N bonds. Indeed, we succeeded to optimize two non-classical $^3$MC states (herein denoted as $^3$MC$_1$ and $^3$MC$_2$) by manually enlarging both Ir-N bond distances in a symmet-ric and in an antisymmetric fashion; respectively. Figure 2a-b displays their main geometrical features; i.e., enlarged axial Ir-N bond distances of 2.45 and 2.55 Å for $^3MC_1$ and the elongation of a single axial Ir-N bond (2.57 Å) for $^3MC_2$; respectively. Importantly, the pseudo-octahedral disposition at both the $^3MC_1$ and $^3MC_2$ geometries is highly distorted. As none of the Ir-N bonds is fully broken, a trigonal bipyramid disposition is never reached for these non-classical $^3MC$ states of **2** (see Figure 2a-b). Similar non-classical $^3MC$ states were also found for heteroleptic tris(bidentate) Ir(III) complexes bearing two Ir-N bonds in trans position,[22] but also for $[Ru(bpy)_3]^{2+}$.[23] In the case of $^3MC_1$ (see Figure 2a), its LUMO NTO clearly displays a $d_{z^2}$ character, as the enlargement of both Ir-N trans positions leads to a stabilization of the latter virtual orbital. Conversely, for $^3MC_2$ (see Figure 2b), its LUMO is predominantly of $d_{x^2-y^2}$-character. Finally, a classical $^3MC$ state could also be optimized for *mer*-Ir(ppy)$_3$, see $^3MC_3$ in Figure 2c. $^3MC_3$ displays a fully rupture of the Ir-N bond (3.39 Å) which is located in *trans* position to the Ir-C bond, and thus a typical trigonal bipyramid arrangement is attained at $^3MC_3$. Concomitantly, both Ir-N bond distances are shortened (2.07 and 2.08 Å) at this geometry. Consequently, its LUMO is predominantly of $d_{x^2-y^2}$-character (see Figure 2c).

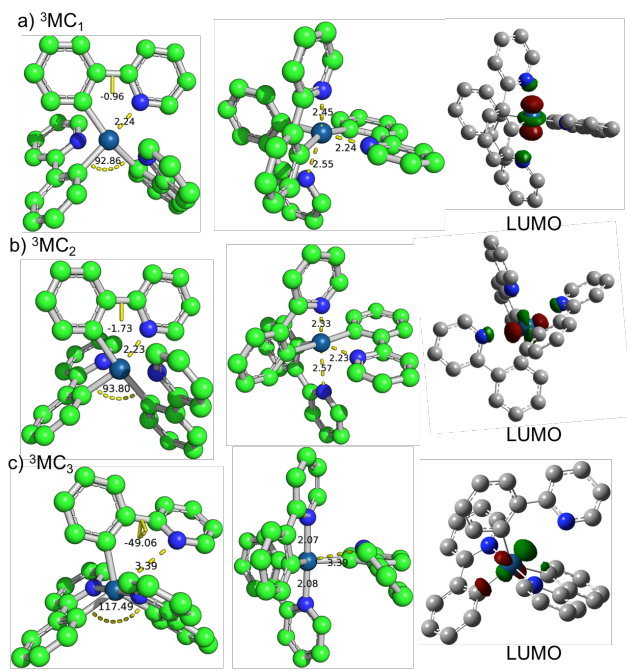

**Figure 2.** Relevant geometrical parameters at the $^3MC_1$ (a), $^3MC_2$ (b), and $^3MC_1$ (c) optimized geometries of 2. Cenital (left) and frontal (middle) views of the $^3MC$ minima along with their computed LUMO NTO (right).

We computed the relevant stationary points along the three different $^3MLCT \rightarrow {}^3MC_{1-3} \rightarrow {}^1GS$ pathways for **2**, and they are schematically depicted in Figure 3. Importantly, all these deactivation pathways possess an energetically accessible $^1GS/^3MC$ MECP that suppresses photoluminescence. In view of the PWPB95 activation barriers (17.58, 16.25 and 7.71 kcal/mol, see Table 1 and Figure 3) for the $^3MLCT \rightarrow {}^3MC_1$, $^3MLCT \rightarrow {}^3MC_2$ and $^3MLCT \rightarrow {}^3MC_3$ pathways, respectively; only the latter channel is readily available to be populated at RT. Hence, the $^3MLCT \rightarrow {}^3MC_1$ (symmetric stretching) and $^3MLCT \rightarrow {}^3MC_2$ (antisymmetric stretching) nonradiative channels may only be populated at temperatures above RT. Comparing the results for **1** and **2**, we can now explain why the *mer*-isomer possesses a one order of magnitude larger $k_{nr}$ value at RT than its *fac*-counterpart.[17] The limiting barrier for its lower-lying temperature-dependent channel, i.e., $^3MLCT \rightarrow {}^3MC_3 \rightarrow {}^1GS$, is ca. 2 kcal/mol smaller than in the *fac*-isomer (compare 9.6 kcal/mol for **1** vS 7.7 kcal/mol for **2** in Table 1). As recently investigated,[21] such a moderate effect in the magnitude of the activation barrier has a tremendous effect on the temperature-dependent nonradiative rates, i.e., $k_{nr}(T)$, since there is an exponential relationship between both magnitudes. Thus, only in the case of **2** the $k_{nr}(T)$ channel is competitive to $k_r$ and $k_{ISC}$ at RT. Consequently, much smaller radiative efficiencies are observed for **2** in comparison with **1**. In addition, we can also provide some insights into the easy of the *mer*-to-*fac* photoisomerization process, which is irreversible upon continuous irradiation at RT.[17,24] At the $^3MC_1$ geometry of **2** (see Figure 2c), the monocoordinated ligand can rotate itself without steric hindrance between the ligands to yield the *fac*-isomer. Note that a simple single rotation at the $^3MC_3$ geometry, which is the most easily populated $^3MC$ structure, will only regenerate the *mer*-isomer. Since the $^3MC_1$ geometry is hardly populated at RT, we therefore conclude that the photoisomerization mechanisms in **2** are likely more complex than *via* a single ligand rotation. Possible mechanistic scenarios arise from concerted motions between the different ligands. In this regard, future work should be devoted to put the photoisomerization mechanisms on firmer grounds. Indeed, experimentally, this photoisomerization reaction is accelerated in coordinating solvents,[17] which could potentially be explained by a solvent-assisted mechanism also involving the dissociation-association of a solvent molecule to photoisomerization active $^3MC$ intermediates. Finally, since the *fac*-isomer is the thermodynamic product and furthermore, its $^3MC$ geometry cannot easily be populated at RT, the *mer*-to-*fac* photoisomerization reaction is, in practical terms, irreversible. The population of the $^3MC$ states is also related to phosphor degradation upon PhOLED operation. In this regard, in view of the computed evidences, i.e., the $^3MC$ state of **2** being more easily accessible than that of **1**, we predict that OLEDs devices using **2** as a dopant would be less efficient and also more prone to degradation than those using **1** as a dopant.

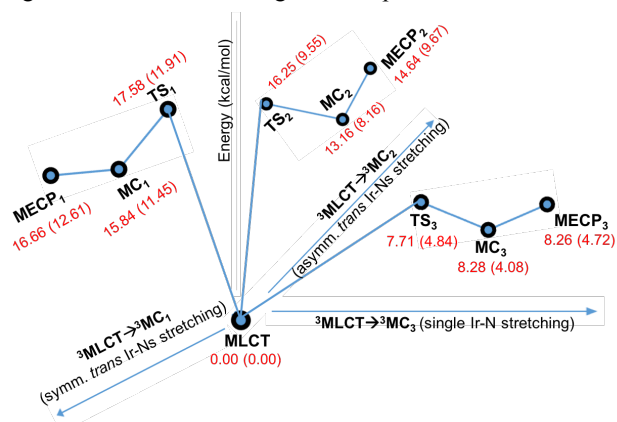

**Figure 3.** PWPB95/6-31G* (B3LYP values between parentheses) energetic profile of the temperature-dependent

nonradiative pathways of 2. The reference is the $^3$MLCT emissive state.

**Table 1. PWPB95 (B3LYP) Relative Energy Gaps (in kcal/mol) Between the Stationary Points along the $^3$MLCT→$^3$MC→$^1$GS Channels and the $^3$MLCT state for 1-6.**

| Complex | $^3$MLCT | TS | $^3$MC | MECP |
|---|---|---|---|---|
| **1** | 0.00 (0.00) | 9.60 (6.61) | 10.52 (5.13) | 10.64 (5.76) |
| **2** ($^3$MC$_1$ channel) | 0.00 (0.00) | 17.58 (11.91) | 15.84 (11.45) | 16.66 (12.61) |
| **2** ($^3$MC$_2$ channel) | 0.00 (0.00) | 16.25 (9.55) | 13.16 (8.16) | 14.64 (9.67) |
| **2** ($^3$MC$_3$ channel) | 0.00 (0.00) | 7.71 (4.84) | 8.28 (4.08) | 8.26 (4.72) |
| **3** | 0.00 (0.00) | - | -18.39 (-7.07) | -18.02 (-6.19) |
| **4** ($^3$MC$_1$ channel) | 0.00 (0.00) | 0.91 (-2.86) | 0.61 (-3.09) | 0.26 (-2.01) |
| **4** ($^3$MC$_2$ channel) | 0.00 (0.00) | - | -4.15 (-8.25) | -4.04 (-7.46) |
| **4** ($^3$MC$_3$ channel) | 0.00 (0.00) | - | -2.64 (-7.02) | -2.42 (-6.13) |
| **5** | 0.00 (0.00) | 4.13 (8.68) | 4.70 (8.13) | 4.70 (9.07) |
| **6** | 0.00 (0.00) | 5.24 (4.96) | 4.24 (3.50) | 3.17 (4.56) |

*fac-/mer-*Ir(ppz)$_3$ **(3/4):** Complexes **3** and **4** were selected in view of their available experimental photophysical properties.[13,25,26] Importantly, PhOLEDs degradation studies using **3-4** as dopants were also performed.[18] The temperature-dependent nonradiative pathway of *fac*-Ir(ppz)$_3$ (**3**) is well described, and it involves the classical trigonal bipyramid-like $^3$MC state.[15] The relevant geometrical features at the $^3$MC minimum along with its LUMO are shown in Figure 4a. It displays the fully rupture of one axial Ir-N bond (3.69 Å), whilst its LUMO mainly involves the $d_{x^2-y^2}$-like $e_g$ orbital. In agreement with previous reports[15,27] all attempts to optimize a transition state on the $^3$MLCT→$^3$MC deactivation channel were unsuccessful. Therefore, this channel is likely barrierless, and its rate limiting step is the small barrier to surpass the $^1$GS/$^3$MC MECP geometry (ca. 0.4 kcal/mol, see Table 1). In addition, all the attempts to optimize other non-classical $^3$MC states for **3** were unsuccessful. All these evidences perfectly match the experimental evidences. Indeed, a huge decrease of the photoluminescence lifetimes was measured from 150 K to 200 K,[13] leading to a complete disappearance of photoluminescence at RT, which is due to the presence of this low-activation barrier $k_{nr}(T)$ channel.

Let's now turn the discussion to *mer*-Ir(ppz)$_3$ (**4**). Complex **4** has not been investigated so far. Analogously to **2**, three different $^3$MLCT→$^3$MC$_{1-3}$→$^1$GS pathways were found for **4**. The main geometrical features at the $^3$MC$_{1-3}$ minima along with their respective LUMO NTOs are shown in Figure 4b. The $^3$MC$_{1-3}$ characteristics are common for **2** and **4**; and thus, they are not further discussed herein. We computed the relevant stationary points along these three pathways, and the relative energetic gaps between all the stationary points and the $^3$MLCT state are collected in Table 1. According to the PWPB95 calculations, the $^3$MC$_2$ (asymmetrically-stretched *trans* Ir-Ns bonds) and $^3$MC$_3$ (singly stretched Ir-N bond) states can be populated in a barrierless manner from the $^3$MLCT state. In addition, the activation barrier to populate the $^3$MC$_1$ state is very small (0.91 kcal/mol, see Table 1). Therefore, all these nonradiative channels can easily be populated at RT and are responsible for the non-emissive nature of **4** at this temperature. We remark that the excited state reactivity of **4** is slightly different from that of *mer*-Ir(ppy)$_3$, since the population of the $^3$MLCT→$^3$MC$_1$→$^1$GS and $^3$MLCT→$^3$MC$_2$→$^1$GS channels is not feasible at RT for the latter compound.

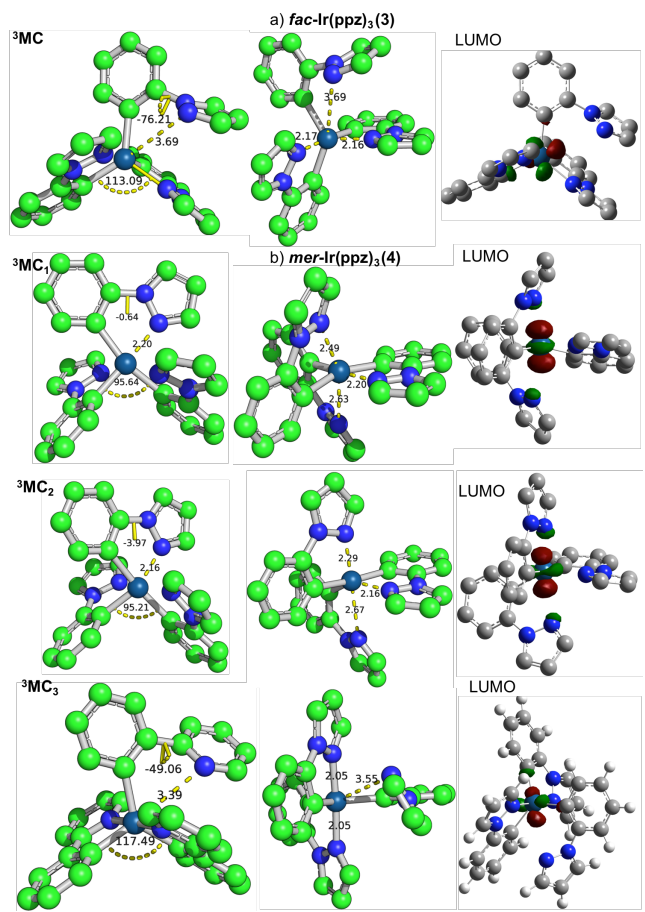

**Figure 4.** Relevant geometrical parameters at the $^3$MC optimized geometries of 3 (a) and 4 (b). Cenital (left) and frontal (middle) views of the $^3$MC minima along with their computed LUMO NTO (right).

Comparing **3** and **4**, ca. 2-3 times larger $k_{nr}$ values were measured for the *mer*-isomer at RT.[25] The fact that at RT, three different nonradiative channels are fully operative for **4** whilst only one is availbale for **3**, likely explain these differences. In addition, rapid *mer*-to-*fac* photoisomerization was also observed for Ir(ppz)$_3$.[26] Finally, *in situ* observation of degradation by ligand substitution was observed in PhOLEDs devices bearing both **3** and **4** as dopant molecules.[18] Both the *fac*- and *mer*-isomers undergo rapid dissociation of a ppz ligand and then coordinate a host molecule upon PhOLED operation, which is in accordance to their computed energetic profiles of formation of their ³MC states. As a result, an operational lifetime of T80<40min (where TX is the time under constant current operation in which the luminance decreases to X% of its initial value) is obtained for the *fac*-based device. The *mer*-based device shows even faster rates of degradation,[18] in agreement with the higher lability predicted by our calculations.

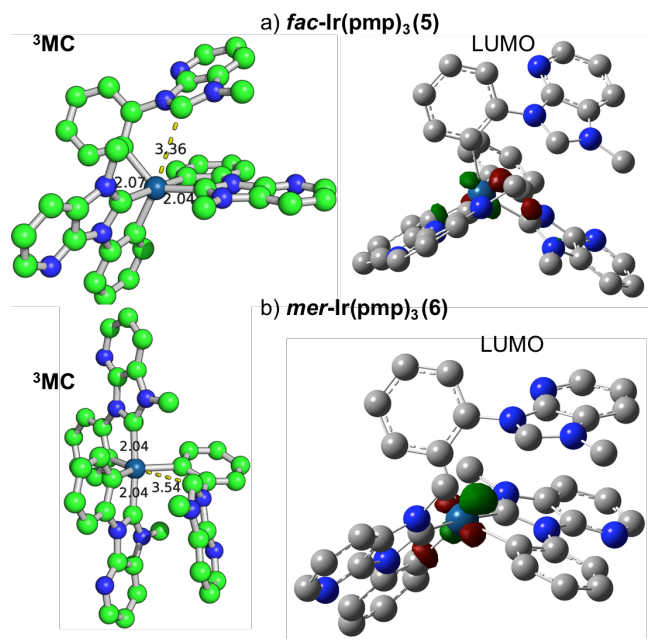

**Figure 5.** a) Relevant geometrical parameters at the ³MC optimized geometries of 5 (a) and 6 (b). Frontal (left) view of the ³MC minima along with their computed LUMO NTO (right).

***fac*-/*mer*-Ir(pmp)₃ (5/6):** The NHC-based Ir(III) complexes **5** and **6** were chosen to rationalize the origins of the subtle differences between the excited state reactivity of their *mer*- and *fac*-isomers. In addition, complete experimental data including PhOLEDs´ performance studies are available for both **5-6**.[20] Up to date, the PES of complexes **5-6** were not computationally investigated. Note that these complexes do not bear weak Ir-N bonds, such as in the case of complexes **1-4**, but strong Ir-carbene bonds instead. This leads to enormous differences in their excited state reactivity. Still, as recently reported for other NHC-based Ir(III)-complexes, the breaking of the Ir-carbene bond is observed at their ³MC state.[15] We deeply explored the PES of **5-6**, and only one ³MC state was found for each complex. In Figure 5a-b are shown the geometrical features of the ³MC state of the *fac*- and *mer*-isomer; respectively, along with their LUMO NTOs. Both the ³MC state of **5** and **6** display elongated Ir-carbene distances (3.36 and 3.54 Å, respectively), in agreement with the previous findings. In the case of **6**, the broken Ir-carbene bond is *trans* to a cyclometalating position. Further attempts to optimize other nonclassical ³MC states were unsuccessful. Whilst this is not surprising in the case of the *fac*-isomer, we believe that the presence of the strong Ir-carbene bonds, strongly destabilize the ³MC$_{1-2}$-like states for the *mer*-isomer. Thus, in conclusion, rather similar ³MLCT→³MC→¹GS nonradiative channels are obtained for both isomers. We also computed all the relevant stationary points along these pathways (see the energetic gaps in Table 1). According to the PWPB95 calculations, **6** possesses a similar activation barrier (5.24 kcal/mol) than that of **5** (4.13 kcal/mol) to populate the ³MC state. These facts are in agreement with the experimental observations of very similar emissive performances,[20] as the photoluminescence quantum yield of **6** (78%) is slightly larger than that of **5** (76%) at 295 K, and thus within the experimental error. Finally, both complexes are highly emissive (95%) at 77K. The NHC-based complexes are known to hardly isomerize photochemically. This is likely connected to the lack of an active ³MC intermediate both for the *mer*- and *fac*-isomers. A recent computational study on similar NHC-complexes disclosed an acidic catalyzed mechanism for *mer*-to-*fac* isomerization involving Ir-phenyl bond dissociation.[28] Finally, PhOLEDs devices using the *mer*-isomer are also more efficient,[20] which is probably due to different degrees of the emitter aggregation quenching between both isomers, but also due to the larger radiative efficiencies of **6**.

**Conclusions**

The different excited state reactivity between the *fac*- and *mer*-isomers of tris(bidentate) Ir(III) complexes have remained, up to date, not completely understood. While several authors hypothesized the presence of different nonradiative channels for *mer*- and *fac*-isomers,[17,19,26] a compelling proof of their actual mechanisms is missing in the literature. In this manuscript we explore in a series of tris(bidentate) Ir(III) complexes the triplet excited state PES of their stereoisomers. In the common C^N cyclometalated complexes, *mer*-isomers are often more labile and less emissive than *fac*-isomers. According to our calculations this is due to: i) the presence of three different low-lying ³MC deactivation pathways (whilst only one remains active for the *fac*-isomers); along ii) with the smaller activation barriers to populate these ³MC states in the case of *mer*-derivatives. Conversely, the presence of strong Ir-carbene bonds in NHC-based complexes leads to the disappearance of some ³MLCT→³MC deactivation channels in their *mer*-isomers. Consequently, only one ³MC deactivation pathway remains active for both their *fac*- and *mer*-isomers, and thus leading to similar excited state reactivity for both isomers. As the population of ³MC states also plays a crucial role on the photoisomerization reactions and on phosphor degradation upon PhOLED operation, this study provides important insights for the design of phosphors with improved PhOLEDs' operational stability. Our computational investigations also highlight the fact that the triplet excited state PES of tris(bidentate) Ir(III) complexes might be more complicated than generally believed in the community (i.e., not only a single ³MC state involved in the temperature-dependent nonradiative pathways), and thus, we generally recommend for future investigations to fully explore the lowest adiabatic triplet PES of these complexes. This should clearly be further explored in other type of architectures, such as e.g., heteroleptic Ir(III) complexes, where examples of the influence of stereoisomerism on OLEDs´ efficiency were reported.[29] Finally, from a computational viewpoint, and in agreement with previous reports,[21,30] the use of the double-hybrid PWPB95 functional is recommended for future explorations of 5d organometallic complexes, especially to get accurate activation energies. Future work should be devoted to study with multiscale modelling approaches the excited state reactivity of the different isomers in the solid state or when used as dopants in organic semiconductor layers.

## EXPERIMENTAL SECTION

**Computational details:** All calculations are based on density functional theory (DFT). The geometries of the singlet ground state ($^1$GS), the triplet $^3$MLCT and $^3$MC states as well as of the transition states (TS) were optimized for **1-6** using the hybrid functional B3LYP[31,32] in combination with the 6-31G* basis set for all atoms. Relativistic effects were considered for the Ir atom by using the ECP-60-mwb pseudopotential.[33] The initial guess geometries for the $^3$MC optimizations were obtained from TD-DFT preoptimizations of the state of interest and/or by manually modified guess structures. The minimum energy crossing points (MECP) between the $^1$GS and the $^3$MC potential surfaces were optimized using Harvey´s algorithm,[34] as implemented in the ORCA software;[35] with the B3LYP functional in combination with the same pseudopotential for Ir and the def2-svp basis set for the other atoms. The Hessian was computed at the same level of theory to confirm the nature of the stationary points. Intrinsic reaction coordinate (IRC) calculations were performed to connect all relevant stationary points. Finally, to get accurate activation energies, single-point calculations were performed with the ORCA software using the double-hybrid PWPB95 functional[36] and the def2-SVP basis set (ECP-60-mwb pseudopotential for Ir). As previously reported, the latter functional, which incorporates only the opposite-spin correlation component, outperforms the rest of functionals for the thermochemistry and kinetics of 5d transition metal complexes.[29] Apart from the MECP optimization and the single point PWPB95 calculations, all the calculations were carried out with the Gaussian09 program package.[37] In order to indistinguishably assign the character of the $^3$MC states, natural transition orbitals (NTOs) were generated from the TD-B3LYP calculations at the $^3$MC optimized geometries.

## ASSOCIATED CONTENT

### Supporting Information

The Cartesian coordinates of all the stationary points along the $^3$MLCT→$^3$MC→$^1$GS deactivation channels for **1-6**.

## AUTHOR INFORMATION

### Corresponding Author

* daniel.escudero@univ-nantes.fr

### ORCID

Daniel Escudero: 0000-0002-1777-8578

### Present address

Daniel Escudero: Department of Chemistry, KU Leuven, Celestijnenlaan 200F, B-3001 Leuven, Belgium.

### Notes

The authors declare no competing financial interest.

## ACKNOWLEDGMENTS

DE thanks funding from the European Union's Horizon 2020 research and innovation programme under the Marie Sklodowska-Curie grant agreement No 700961.

TOC

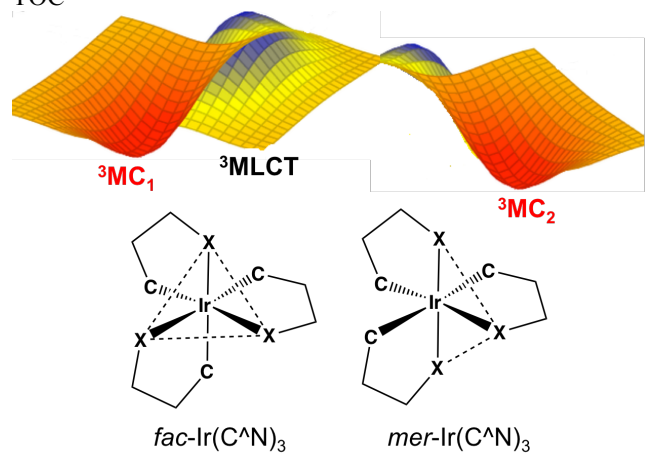

The triplet potential energy surfaces of *mer*- and *fac*-isomers are deeply explored. The presence of different $^3$MC states strongly influences their excited state reactivity, especially for the *mer*-isomers.